\documentclass{article}
\usepackage{spconf,amsmath,graphicx}
\usepackage{hyperref}
\usepackage{color}
\usepackage{amssymb}
\usepackage{booktabs}
\usepackage{setspace}

\title{Xiaoicesing 2: A High-Fidelity Singing Voice Synthesizer Based on Generative Adversarial Network}

\name{$^*${Chunhui Wang}$^1$, $^*${Chang Zeng}$^{2,3}$, Xing He$^1$}
\address{$^1$Beijing Bombax XiaoIce Technology Co., Ltd, China \\ $^2$National Institute of Informatics, Japan $^3$SOKENDAI, Japan}

\begin{document}
\ninept
\maketitle
\renewcommand{\thefootnote}{\fnsymbol{footnote}}
\footnotetext[1]{These authors contributed equally to this work.}
\begin{abstract}
\vspace{0mm}
XiaoiceSing is a singing voice synthesis (SVS) system that aims at generating $48$kHz singing voices. However, the mel-spectrogram generated by it is over-smoothing in middle- and high-frequency areas due to no special design for modeling the details of these parts. In this paper, we propose XiaoiceSing2, which can generate the details of middle- and high-frequency parts to better construct the full-band mel-spectrogram. Specifically, in order to alleviate this problem, XiaoiceSing2 adopts a generative adversarial network (GAN), which consists of a FastSpeech-based generator and a multi-band discriminator. We improve the feed-forward Transformer (FFT) block by adding multiple residual convolutional blocks in parallel with the self-attention block to balance the local and global features. The multi-band discriminator contains three sub-discriminators responsible for low-, middle-, and high-frequency parts of the mel-spectrogram, respectively. Each sub-discriminator is composed of several segment discriminators (SD) and detail discriminators (DD) to distinguish the audio from different aspects. The experiment on our internal $48$kHz singing voice dataset shows XiaoiceSing2 significantly improves the quality of the singing voice over XiaoiceSing.
\end{abstract}
\begin{keywords}
Singing voice synthesis, feed-forward transformer, generative adversarial network
\end{keywords}

\section{Introduction}
\vspace{-2mm}
\label{sec:intro}
Recently neural network for singing voice synthesis \cite{48khz1,lee-koreasvs,xiaoicesing,bytesing} has attracted a lot of attention since deep learning has achieved great gain in text-to-speech (TTS) task \cite{tacotron2, fastspeech, fastspeech2} which has a similar pipeline to SVS. Some studies \cite{48khz1,xiaoicesing,48khz2,48khz3,48khz4,hifisinger} reported the promising results of the proposed models on synthesizing high-fidelity $48$kHz singing voices. For instance, XiaoiceSing \cite{xiaoicesing} modified the architecture of FastSpeech \cite{fastspeech} to adapt the task of high-fidelity SVS and it was combined with WORLD \cite{world} vocoder to generate $48$kHz singing voices. HifiSinger \cite{hifisinger} utilized a sub-frequency GAN in the acoustic model and a multi-length GAN in the vocoder to better reconstruct the high-fidelity singing voices.

However, due to no special design for generating the full-band mel-spectrogram, these studies work not well in high-fidelity SVS scenarios in which middle- and high-frequency parts possess stronger emotion and expressiveness. Further, this over-smoothing problem in the generated mel-spectrogram results in the vocoder failing to reconstruct the high-fidelity waveform from it owing to its low-quality \cite{hifisinger,singgan,revisting-oversmooth}.

In order to solve the over-smoothing problem of middle- and high-frequency areas, we present a novel high-fidelity singing voice synthesizer XiaoiceSing2 based on a generative adversarial network \cite{gan} to generate a more realistic mel-spectrogram since GAN can theoretically approximate the real data distribution via the adversarial training. The proposed XiaoiceSing2 is composed of a FastSpeech-based generator and a multi-band discriminator. For the generator, we follow the design of XiaoiceSing \cite{xiaoicesing} but improve the feed-forward Transformer (FFT) block \cite{s2sfft} of it by adding multiple residual convolutional blocks in parallel with the multi-head self-attention (MHSA) block \cite{attention-all-you-need} to balance the local and global features. Because we argue that the global features generated by the MHSA block are prone to being over-smoothing for the middle- and high-frequency parts, which can be alleviated by introducing local information from the multiple residual convolutional blocks. 

As for the multi-band discriminator, similar to HiFiSinger \cite{hifisinger}, it consists of three sub-discriminators responsible for low-, middle-, and high-frequency parts of the mel-spectrogram, respectively. Moreover, each sub-discriminator contains several segment discriminators (SD) and detail discriminators (DD) for distinguishing the mel-spectrogram from the segments with different window lengths and local time-frequency patterns, respectively. The segment discriminators are able to cover the different levels of long-term dependencies by applying multiple windows with different lengths on the mel-spectrogram to increase the capability of the discriminator. Similar to PatchGAN \cite{patchgan1,patchgan2,patchgan3}, the detail discriminator divides the mel-spectrogram into multiple time-frequency patches so that it can pay more attention to the middle- and high-frequency regions and the generator also benefits from the stronger discriminator to produce a more realistic mel-spectrogram. 

In the experiment, XiaoiceSing2 is combined with a high-fidelity vocoder HiFi-WaveGAN \cite{hwg} which is designed to reconstruct the $48$kHz waveform and the result shows XiaoiceSing2 significantly improves the quality of the singing voice over XiaoiceSing in term of mean opinion score (MOS) metric. We also make a comparative study of the middle- and high-frequency areas generated by XiaoiceSing2 and XiaoiceSing via visualizing the mel-spectrogram. Besides, an ablation study is conducted to show the contribution of the proposed components.

The rest of this paper is organized as below. Section \ref{sec:method} illustrates the detailed architecture of XiaoiceSing2 including the generator and discriminator. The experimental settings including the dataset, baseline system, and training methodology are shown in Section \ref{sec:exp}. In addition, the MOS test result and an ablation study are also reported in this section. Finally, we conclude this paper in Section \ref{sec:con}. The audio samples generated by XiaoiceSing2 can be found at \href{https://wavelandspeech.github.io/xiaoice2}{\texttt{https://wavelandspeech.github.io/xiaoice2}}

\begin{figure*}[t]
\begin{minipage}[b]{0.23\linewidth}
  \centering
  \centerline{\includegraphics[width=4cm]{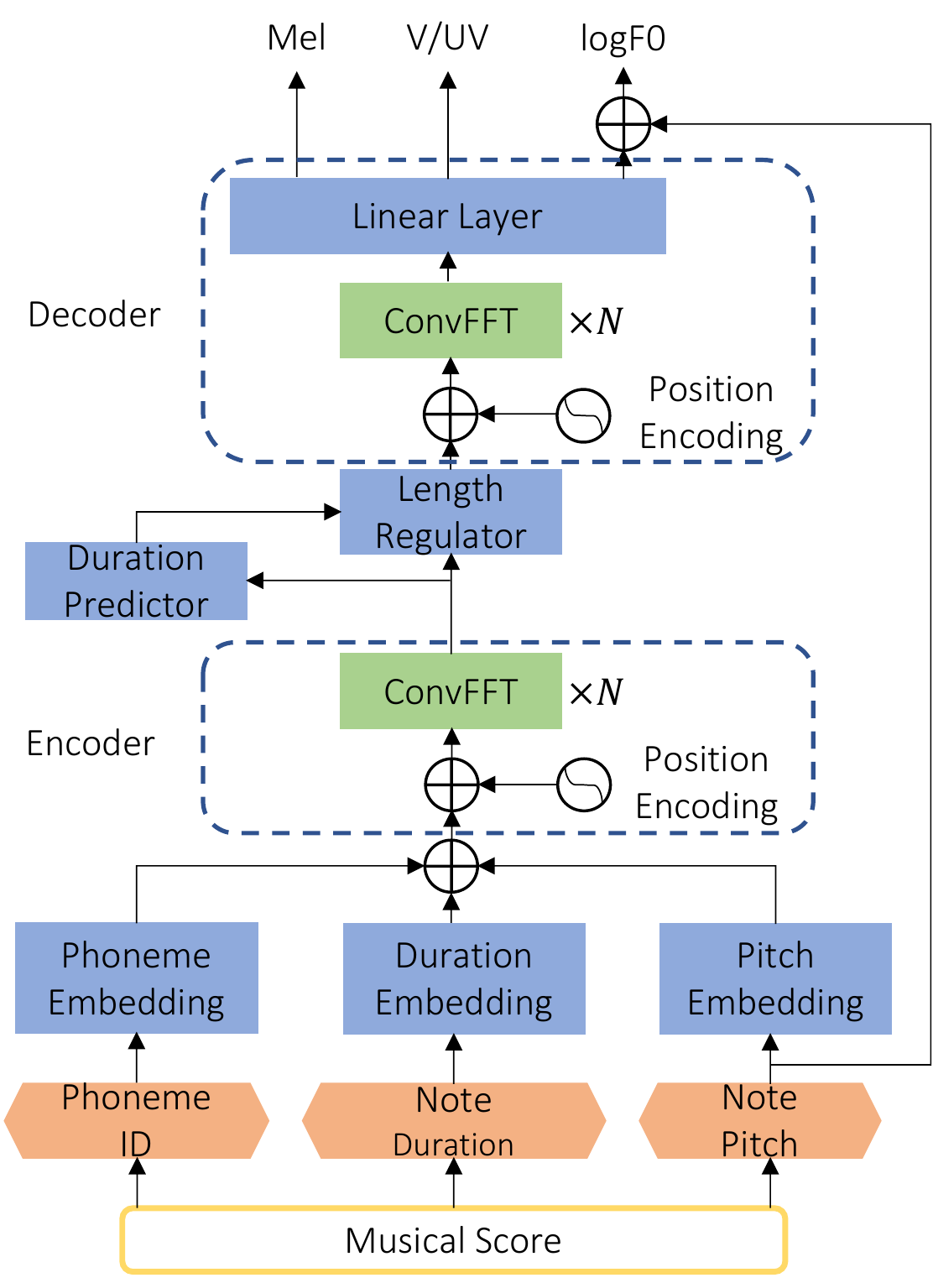}}
  \centerline{(a) Generator}\medskip
\end{minipage}
\begin{minipage}[b]{0.4\linewidth}
  \centering
  \centerline{\includegraphics[width=5cm]{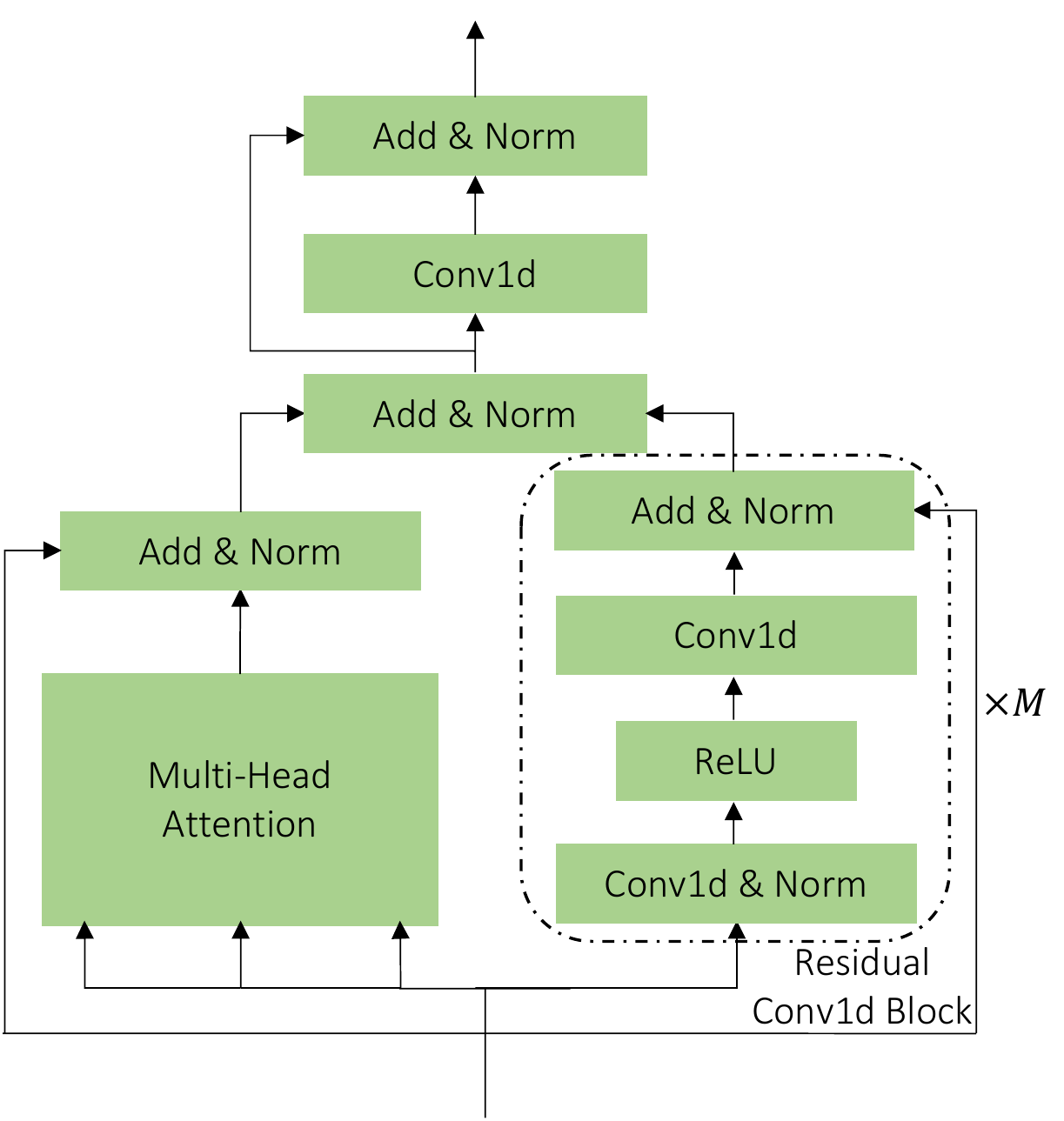}}
  \centerline{(b) ConvFFT block}\medskip
\end{minipage}
\begin{minipage}[b]{0.32\linewidth}
  \centering
  \centerline{\includegraphics[width=7cm]{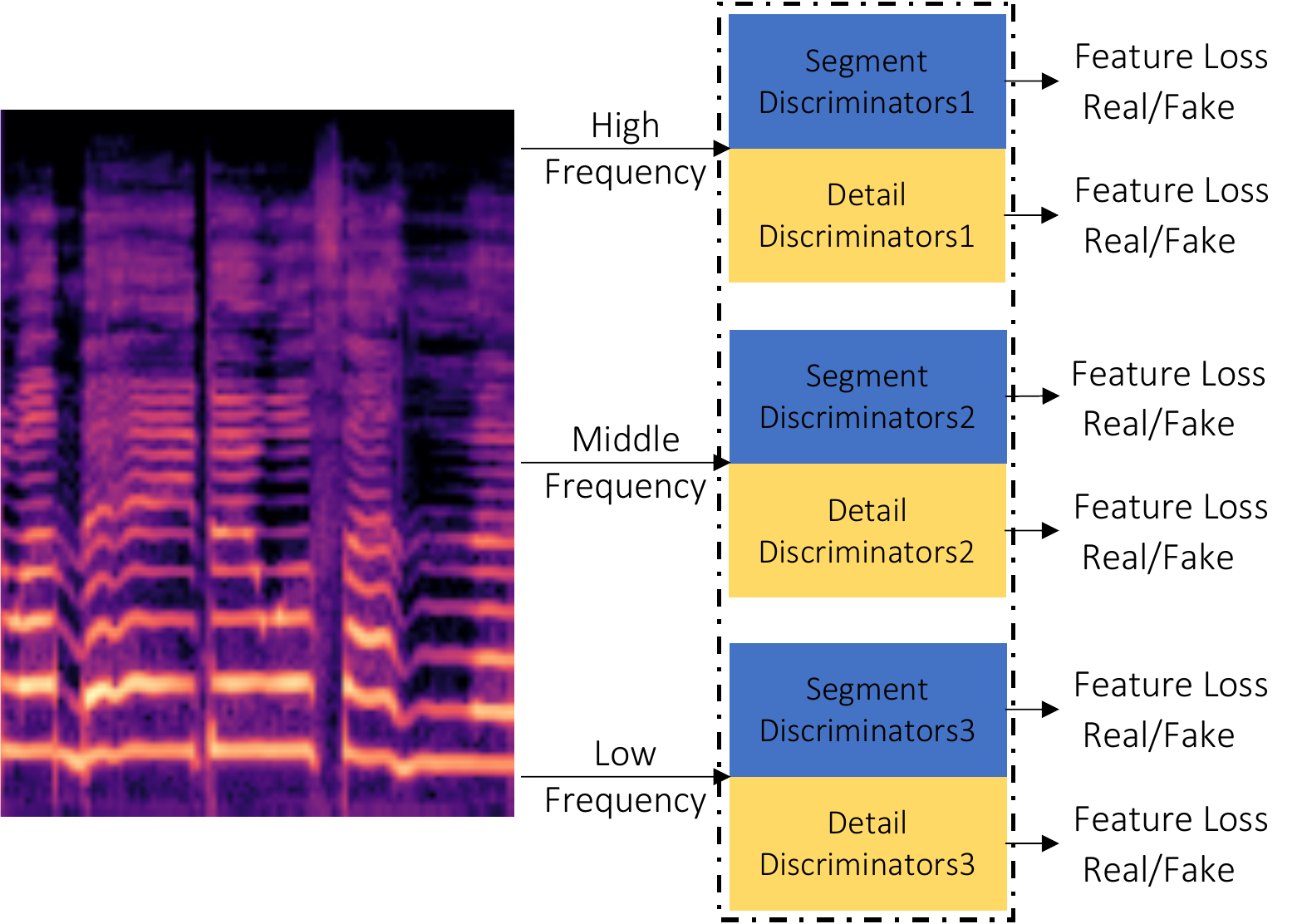}}
  \centerline{(c) Multi-band discriminator}\medskip
\end{minipage}
\caption{The architecture of XiaoiceSing2. (a). The improved feed-forward Transformer. (b). Feed-forward Transformer with parallel residual convolutional block. (c). Multi-band discriminator, consisting of three sub-discriminators, and each contains several segment discriminators and detail discriminators.}
\label{fig:architecture}
\end{figure*}

\section{Proposed Method}
\label{sec:method}
\vspace{0mm}
In order to generate the mel-spectrogram with more fine-grained middle- and high-frequency parts, we adopt an adversarially training strategy to optimize XiaoiceSing2, which is different from XiaoiceSing that directly trains the feed-forward Transformer \cite{s2sfft} as the acoustic model. The generator of XiaoiceSing2 uses the novel ConvFFT blocks to efficiently leverage the local and global information as Figure \ref{fig:architecture}(a) shows. Additionally, since the generator can benefit from a powerful discriminator, we use three sub-discriminators to distinguish the low-, middle-, and high-frequency parts of the mel-spectrogram. And each sub-discriminator employs several segment and detail discriminators to identify the mel-spectrogram from different aspects.

\subsection{Generator}
\label{subsec:ge}
The input to the generator is the musical score consisting of lyrics, note duration sequence, and note pitch sequence. The lyrics are converted to phoneme sequences by the grapheme-to-phoneme (G2P) tool \cite{g2p}. All sequences are transformed into their own embedding spaces by the corresponding embedding modules and these embedding sequences are concatenated as shown in Figure \ref{fig:architecture}(a).

As for the architecture of the generator, XiaoiceSing2 follows XiaoiceSing which is a FastSpeech-based \cite{fastspeech} acoustic model. The generator can be divided into an encoder, a length regulator with a duration predictor, and a decoder as shown in Figure \ref{fig:architecture}(a). The encoder converts the concatenated sequence into a hidden space which is considered to be shared with the mel-spectrogram \cite{fastspeech, fastspeech2}. Consequently, the output sequence of the encoder can be expanded by the length regulator according to the result of the duration predictor to directly match the length of the target mel-spectrogram. Finally, the expanded sequence is transformed by the decoder to predict the mel-spectrogram, V/UV decision, and the logF0 value. Note there is a residual connection between the input pitch sequence and the predicted logF0 sequence to lower the training difficulty \cite{xiaoicesing}.

In this paper, both the encoder and decoder contain $6$ ConvFFT blocks which are improved from the FFT block used in \cite{xiaoicesing,fastspeech,fastspeech2}. As Figure \ref{fig:architecture}(b) shows, the ConvFFT block incorporates multiple residual convolutional blocks in parallel with the MHSA block since we believe that the over-smoothing problem of middle- and high-frequency areas is intensified if only the global information extracted by MHSA is used. To rectify this problem in the generated mel-spectrogram, the local information is extracted by the stacked residual convolutional blocks which share the same input with MHSA, then it is added to the global information for fusion. In the encoder, each ConvFFT block has $2$ residual convolutional blocks. In the decoder, the number is $5$.

\subsection{Multi-band discriminator}
\label{subsec:dis}
Because the strong discriminator is beneficial to the generator, we utilize three sub-discriminators to work on the low-, middle-, and high-frequency parts of the mel-spectrogram as Figure \ref{fig:architecture}(c) shows. In this paper, the dimension of the mel-spectrogram is $120$ and it is divided into low-frequency ($0$-$60$), middle-frequency ($30$-$90$), and high-frequency ($60$-$120$) parts. Each sub-discriminator has several segment and detail discriminators for identifying the mel-spectrogram from long-term dependencies as well as time-frequency patterns.

\subsubsection{Segment discriminator}
The idea of the SD is similar to the multi-length GAN (ML-GAN) in HiFiSinger \cite{hifisinger}. However, instead of applying ML-GAN on the waveform, we utilize our SD for the mel-spectrogram via randomly clipping the input mel-spectrogram by different window lengths which are set as $[200,400,600,800]$, and the whole segment in this paper. All segment discriminators have the same architecture which is a $10$-layers $1$-dimensional ($1$-d) convolutional neural network (CNN) with $3$ kernel size and $128$-dimensional hidden channel. The $1$-d CNN is able to promote the continuity of the mel-spectrogram produced by the generator along the time axis by distinguishing the long-term dependencies with different lengths. In addition to outputting the real/fake decision, the intermediate feature maps generated by the hidden layers are also collected for calculating the feature loss which is described in Section \ref{subsub:fl}.

\subsubsection{Detail discriminator}
Although segment discriminators promote the continuity of the generated mel-spectrogram, they cannot benefit to generate the high-quality middle- and high-frequency parts which are significant to produce high-fidelity singing voices \cite{singgan}. By taking this into account, we accompany a detail discriminator for each segment discriminator. The motivation for generating more fine-grained middle- and high-frequency areas by using the detail discriminator comes from the PatchGAN \cite{patchgan1,patchgan2,cyclegan} which utilizes a fully convolutional discriminator for generating high-resolution images. 
To be specific, the first $2$-dimensional convolutional layer with $(3,3)$ kernel size upsamples the input mel-spectrogram to $32$ channels. The rest of the network consists of $5$ convolutional layers with $(3,3)$ kernel size and $(2,2)$ dilation for downsampling, and $5$ convolutional layers with $(1,3)$ kernel size for output. These layers are alternately stacked. The outputs from each output layer are collected for calculating the adversarial loss. And the outputs from each downsampling layer are collected for computing the feature loss. Due to the detail discriminator, the mel-spectrogram is divided into multiple time-frequency patches for identifying whether they are real or fake, which is helpful to better construct the middle- and high-frequency parts. 

\subsection{Loss function}
The loss for XiaoiceSing2 is a weighted sum of several loss terms, which is formulated as follows,
\begin{align}
    \label{eq:loss}
    \mathcal{L}_G & = \lambda_1 * \mathcal{L}(G;D) + \lambda_2 * \mathcal{L}_a + \lambda_3 * \mathcal{L}_f, \\
    \mathcal{L}_D & = \mathcal{L}(D;G),
\end{align}
where $\mathcal{L}_G$ denotes the generator loss, $\mathcal{L}{(G;D)}$ is the adversarial loss for the generator, $\mathcal{L}_a$ denotes the acoustic loss similar to the one used in \cite{xiaoicesing}, and $\mathcal{L}_f$ represents the feature loss. While $\mathcal{L}_D$ denotes the discriminator loss, which only possesses the adversarial loss term $\mathcal{L}{(D;G)}$ for the discriminator. As for the weights $\lambda_1$, $\lambda_2$, and $\lambda_3$ in Eq. \ref{eq:loss}, they are set as $0.1$, $1$, and $1$ in this paper, respectively.

\subsubsection{Adversarial loss}
The adversarial loss proposed in LS-GAN \cite{lsgan} is also used in the training stage of XiaoiceSing2. The formula is shown as
\begin{align}
    \label{eq:adv}
    \mathcal{L}_{adv}(G;D) & = \mathbb{E}_{\boldsymbol{z} \sim \mathcal{N}(0,1)}[(1 - D(G(\boldsymbol{z})))^2], \\
    \mathcal{L}_{adv}(D;G) & = \mathbb{E}_{\boldsymbol{x} \sim p_{data}}[(1 - D(\boldsymbol{x}))^2] + \mathbb{E}_{\boldsymbol{z} \sim \mathcal{N}(0,1)}[D(G(\boldsymbol{z}))^2],
\end{align}
where $\boldsymbol{z}$ denotes the random noise and $\boldsymbol{x}$ is the real mel-spectrogram. This format of adversarial loss can avoid the gradient vanishing while training the GAN \cite{lsgan}.

\subsubsection{Acoustic loss}
The acoustic loss is also a weighted sum of the loss terms for the predicted acoustic features, which is shown as
\begin{align}
    \label{eq:acoustic}
    \mathcal{L}_a = \alpha_1 * \mathcal{L}_{mel} + \alpha_2 * \mathcal{L}_{pitch} + \alpha_3 * \mathcal{L}_{V/UV} + \alpha_4 * \mathcal{L}_{dur},
\end{align}
where $\mathcal{L}_{mel}$, $\mathcal{L}_{pitch}$, and $\mathcal{L}_{dur}$ are MSE loss for the mel-spectrogram, pitch, and duration, respectively. While $\mathcal{L}_{V/UV}$ is a binary cross-entropy loss for the V/UV decision. Besides, the weights $\alpha_1$, $\alpha_2$, $\alpha_3$, and $\alpha_4$ are set as $1$, $0.01$, $0.01$, and $0.1$, respectively.

\subsubsection{Feature loss}
\label{subsub:fl}
Feature loss was proposed in \cite{fl1} and was introduced into the speech field in MelGAN \cite{melgan}. The generator can make full use of the information brought by the discriminator via learning from the $L1$ similarity metric between the feature maps of the real and fake data. It can be formulated as
\begin{align}
    \label{eq:fl}
    \mathcal{L}_{f} = \mathbb{E}_{\boldsymbol{z}, \boldsymbol{x}}[\sum_{k=1,2,3} ( & \sum_{i=1}^{L_s}\frac{1}{N_i} ||D_{ks}^i(\boldsymbol{x})-D_{ks}^i(G(\boldsymbol{z}))||_1 \nonumber \\
    & + \sum_{j=1}^{L_d}\frac{1}{N_j}||D_{kd}^j(\boldsymbol{x})-D_{kd}^j(G(\boldsymbol{z}))||_1)], 
\end{align}
where $||\ \cdot \ ||_1$ denotes the $L_1$ distance, $D_{ks}^i(\ \cdot \ )$ and $D_{kd}^j(\ \cdot \ )$ denote the feature maps of $i$-th and $j$-th layer of the $k$-th SD and DD, respectively. $N_i$ and $N_j$ are the numbers of the corresponding feature map. $L_s$ and $L_d$ denote the number of layers of SD and DD, respectively. Since the multi-band discriminator has three sub-discriminator, the feature loss for each sub-discriminator is merged in Eq \ref{eq:fl}.

\section{Experiments}
\label{sec:exp}

\subsection{Dataset}
\label{subsec:data}
We conduct the experiment on our internal singing voice dataset including $6917$ pieces of singing voices from a Mandarin female singer, which is identical to the one used in HiFi-WaveGAN \cite{hwg}. All audios are sampled at $48$kHz. The duration of audio in the dataset ranges from $4$s to $10$s and the total duration is $5$ hours. We transform each audio into the corresponding STFT spectrogram by applying $20$ms window with $5$ms shift. The spectrogram is converted to mel-scale by $120$ filters. The pitch and V/UV decision are extracted by using the Parselmouth \cite{parselmouth} toolkit which is a Python interface to Praat \cite{praat}. As for the division of data, we randomly choose $300$ segments for validation and $300$ segments for testing. The remaining data is used for training. 

\subsection{Baseline system}
\label{subsec:baseline}
Since XiaoiceSing2 is improved based on XiaoiceSing, XiaoiceSing is selected as the baseline system. XiaoiceSing adapts FastSpeech \cite{fastspeech} from TTS to SVS by extending the inputs and outputs of the model. It concatenates the pitch sequence, duration sequence, and phoneme sequence as the input to the model and it outputs the mel-spectrogram, V/UV decision, and logF0 for the vocoder. Besides, it uses a residual connection between input pitch and output logF0 to lower the difficulty of training. We train the model with the same optimization strategy in \cite{xiaoicesing}. As for the vocoder, we utilize the HiFi-WaveGAN \cite{hwg} to generate high-fidelity singing voices for fair comparison because it is designed for the scenario of $48$kHz SVS.

\subsection{Training methodology}
\label{subsec:train_me}
XiaoiceSing2 is trained on $4$ NVIDIA V100 GPUs with $32$ batch size for $300$ epochs until convergence, which costs $24$ hours. We use Adam \cite{adam} optimizer with $0.01$ learning rate, $0.9$ $\beta_1$, $0.98$ $\beta_2$, and $10^{-9}$ $\epsilon$ to train the both generator and discriminator. In addition, we adopt a warmup strategy that is identical to the one in \cite{attention-all-you-need} to adjust the learning rate for better optimization.

\subsection{Subjective evaluation}
\label{subsec:results}
To show the quality of the singing voices generated by XiaoiceSing2, we conduct a subjective evaluation for the real and synthesized audio. $20$ listeners are asked to give their opinion score to $20$ segments in terms of quality and naturalness of the singing voices, which indicates we collect $400$ scores for the ground truth as well as each system. Table \ref{tab:result_mos_svs} summarized the evaluation result. Compared with the MOS of XiaoiceSing, the proposed XiaoiceSing2 significantly outperform it by over $0.93$ in term of the MOS metric. And the number of the fluctuation in the $95$\% confidence interval shows XiaoiceSing2 is much more stable than XiaoiceSing when synthesizing high-fidelity singing voices. In addition, the MOS of XiaoiceSing2 is very close ($-0.04$) to the ground truth from the table, which means our model can synthesize the human-level singing voices in the $48$kHz scenario.

\subsection{Spectrogram analysis}
\label{subsec:analysis}
\begin{figure}[t]
\begin{minipage}[b]{0.5\linewidth}
  \centering
  \centerline{\includegraphics[width=4.5cm]{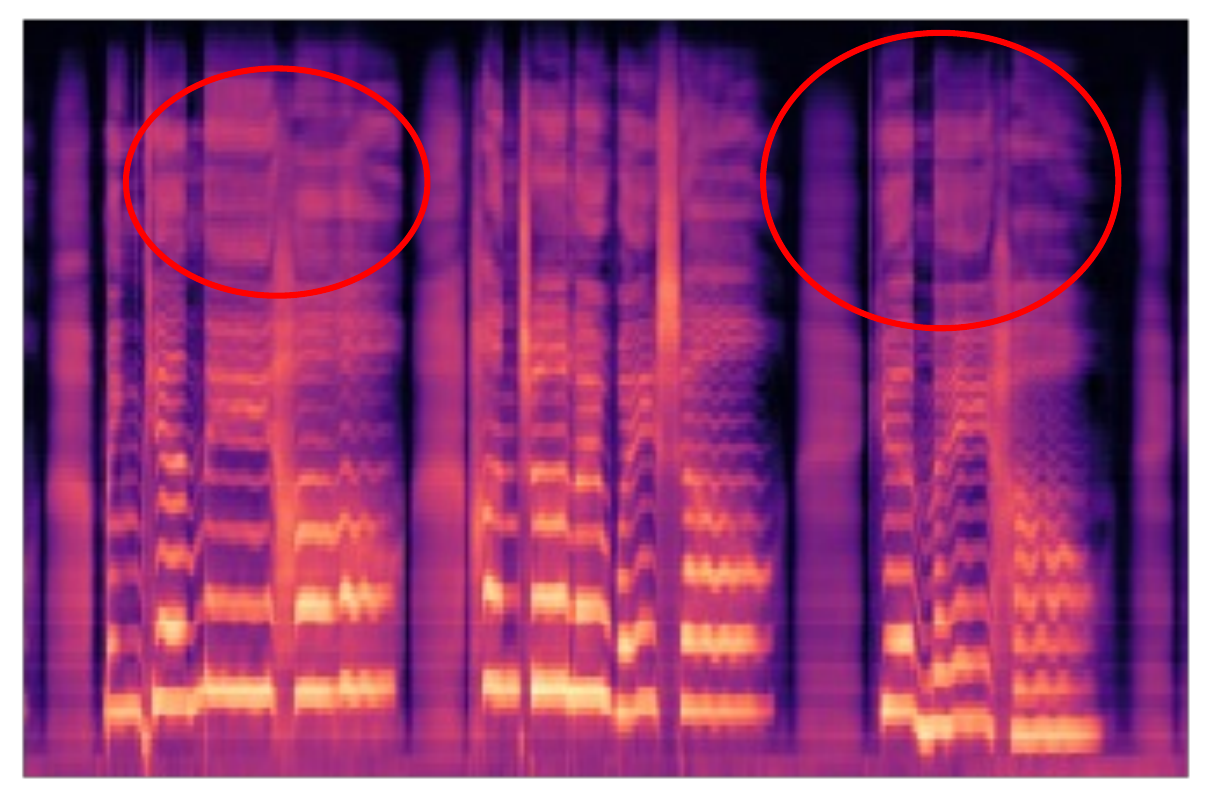}}
  \centerline{(a) XiaoiceSing}\medskip
\end{minipage}
\begin{minipage}[b]{0.48\linewidth}
  \centering
  \centerline{\includegraphics[width=4.5cm]{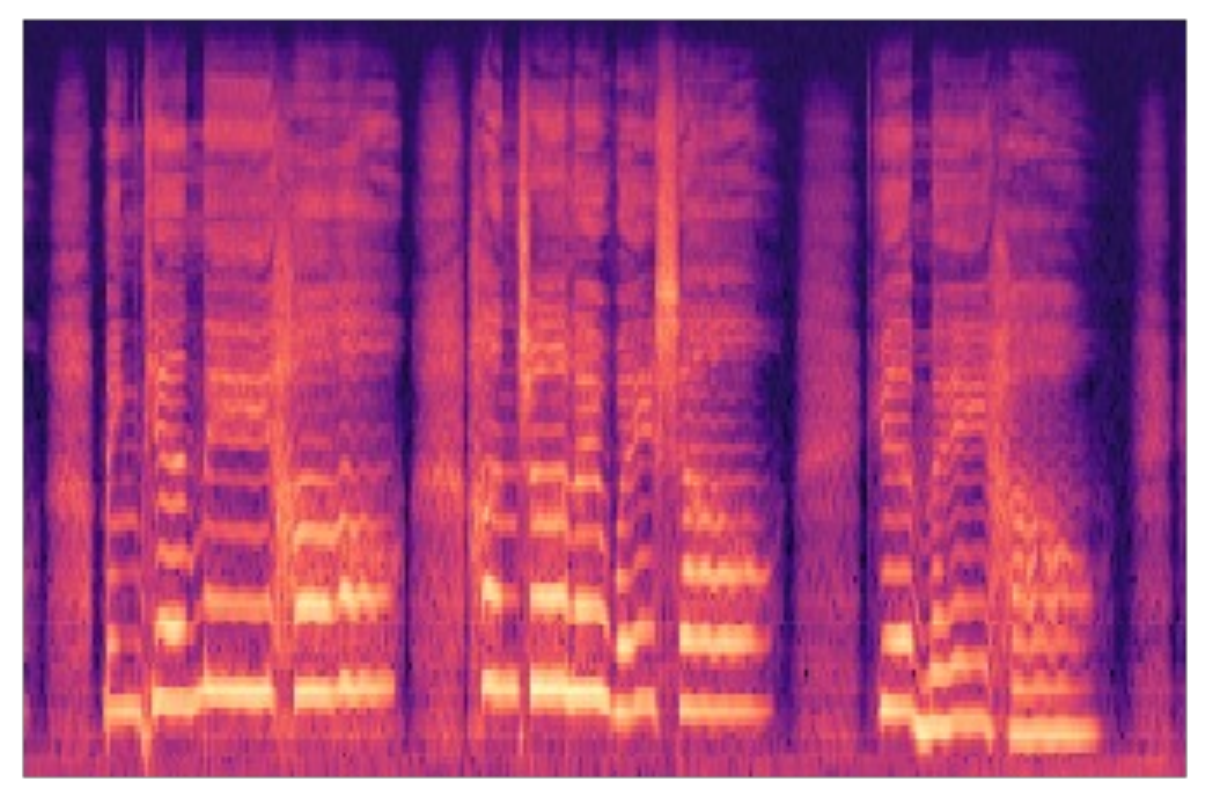}}
  \centerline{(b) XiaoiceSing2}\medskip
\end{minipage}
 \caption{Mel-spectrograms generated by XiaoiceSing and XiaoiceSing2, respectively.}
 \label{fig:mel}
\end{figure}

Although the MOS test result indicates the quality of the singing voices generated by XiaoiceSing2 is much better than it of XiaoiceSing, it is necessary to find some evidence from the generated mel-spectrograms to support this conclusion. As Figure \ref{fig:mel} shows, it seems the left mel-spectrogram generated by XiaoiceSing has more distinct spectral lines compared with the right mel-spectrogram generated by XiaoiceSing2. However, the distinct mel-spectrogram also indicates the severe over-smoothing problem for high-fidelity singing voice generation. Compared with Figure \ref{fig:mel}(a), Figure \ref{fig:mel}(b) obviously reserves more details in the transition regions between the adjacent spectral lines, which demonstrates the over-smoothing problem is alleviated. In addition, the over-smoothing problem in the high-frequency parts circled in Figure \ref{fig:mel}(a) also leads to audible hissing noise in the generated audio.

\setlength{\tabcolsep}{4mm}
\begin{table}[t]
\footnotesize
  \caption{MOS test result with $95$\% confidence interval of the ground truth and different acoustic models for 48kHz singing voice synthesis.}
  \label{tab:result_mos_svs}
  \centering
  \vspace{2mm}
  \begin{tabular}{lc}
    \toprule
    \textbf{Vocoder} & \textbf{MOS}($\uparrow$) \\
    \midrule
    Ground truth                    & $4.27\pm 0.044$          \\
    XiaoiceSing + HiFi-WaveGAN      & $3.30\pm 0.073$                   \\
    XiaoiceSing2 + HiFi-WaveGAN     & $\boldsymbol{4.23}\pm 0.044$         \\
    \bottomrule
  \end{tabular}
\end{table}

\setlength{\tabcolsep}{4mm}
\begin{table}[t]
\footnotesize
  \caption{Ablation study to show the contribution of the proposed components. The XiaoiceSing model of the first line predict MGC and BA rather than the mel-spectrogram. While other models predict the mel-spectrogram.}
  \label{tab:ablation}
  \centering
  \vspace{2mm}
  \begin{tabular}{lc}
    \toprule
    \textbf{Vocoder} & \textbf{MOS}($\uparrow$) \\
    \midrule
    XiaoiceSing + WORLD \cite{xiaoicesing} & $3.39\pm 0.058$   \\
    XiaoiceSing + HiFi-WaveGAN &  $3.30 \pm 0.073$  \\
    (+ConvFFT) + HiFi-WaveGAN  &  $3.33 \pm 0.072$  \\
    (++SD) + HiFi-WaveGAN      &  $4.20 \pm 0.043$   \\
    (+++DD) + HiFi-WaveGAN     &  $4.23 \pm 0.044$\\
    \bottomrule
  \end{tabular}
\end{table}

\subsection{Ablation study}
\label{subsec:abs}
In this paper, we proposed multiple points to improve the quality of high-fidelity singing voices. It is reasonable to figure out the contribution of each proposed component to the quality of the synthesized audio. Therefore, we conduct an ablation study to demonstrate the improvement of each component. As Table \ref{tab:ablation} shows, the first line of it indicates the result of the original XiaoiceSing system described in \cite{xiaoicesing}. Instead of predicting the mel-spectrogram, it generates the mel-generalized cepstrum (MGC) and band aperiodicity (BA) for the WORLD vocoder \cite{world}. The result of it is slightly better than the combination of XiaoiceSing described in Section \ref{subsec:baseline} and HiFi-WaveGAN vocoder \cite{hwg} because the model of the original XiaoiceSing learns more information from the training data. 

When the ConvFFT module is incorporated into the XiaoiceSing as the third line shows, the MOS is boosted by $0.03$, which means even only substituting the FFT in XiaoiceSing with the ConvFFT module, it is helpful to generate a better mel-spectrogram. Based on it, we change the sequence-to-sequence (S2S) model of XiaoiceSing to a GAN-based model by adding all segment discriminators as the fourth line shows. The MOS metric is largely promoted from $3.33$ to $4.20$ as expected, which proves that the GAN-based model has a huge advantage over the S2S model for high-fidelity singing voice synthesis. The last line in the table shows the result of the proposed XiaoiceSing2. By combining the segment and detail discriminators, the MOS is improved by $0.03$ further because of the better construction of the middle- and high-frequency parts.

\section{Conclusion}
\vspace{0mm}
\label{sec:con}
We propose a novel GAN-based acoustic model XiaoiceSing2 for SVS in this paper to relieve the over-smoothing problem in the middle- and high-frequency parts of the mel-spectrogram. In the FastSpeech-based generator, the new ConvFFT block combines the MHSA block and multiple residual convolutional blocks in parallel to couple the global and local information, which is beneficial to generate a more fine-grained mel-spectrogram as shown in the experiment. As for the discriminator, we extend the multi-band discriminator used in HiFiSinger by randomly clipping the mel-spectrogram into several segments so that the discriminator can increase the capability from the different long-term dependencies. Additionally, a detail discriminator accompanying the segment discriminator is used to pay more attention to middle- and high-frequency parts of the mel-spectrogram. The powerful discriminator also forces the generator to produce a more realistic mel-spectrogram. The experimental result on the $48$kHz singing voice dataset proves that XiaoiceSing2 is able to generate high-quality mel-spectrogram, especially in middle- and high-frequency regions.

\section{Future work}
In the future, we will focus on how to efficiently utilize the high-frequency parts of real singing voice data since it occupies only a small part of the training data, which may lead to the synthesizer biases to only learn from low- and middle-frequency of the real data.


\vfill\pagebreak

\bibliographystyle{IEEEbib}
\begin{spacing}{0.5}
\bibliography{strings,refs}
\end{spacing}

\end{document}